%% file: topological-dance_v2.tex
\documentclass[12pt]{article}
\usepackage[latin9]{inputenc}
\usepackage{color}
\usepackage{amsmath}
\usepackage{graphicx}
\PassOptionsToPackage{normalem}{ulem}
\usepackage{ulem}

\makeatletter
\newenvironment{lyxlist}[1]
	{\begin{list}{}
		{\settowidth{\labelwidth}{#1}
		 \setlength{\leftmargin}{\labelwidth}
		 \addtolength{\leftmargin}{\labelsep}
		 }}
	{\end{list}}

\@ifundefined{date}{}{\date{}}

\usepackage{scicite}

\usepackage{times}

\newenvironment{sciabstract}{%
\begin{quote} \bf}{\end{quote}}

\title{Chiral edge waves in a dance-based human topological insulator}

\author
{Matthew Du,$^{1}$ 
Juan B. P\'erez-S\'anchez,$^{1}$
Jorge A. Campos-Gonzalez-Angulo,$^{1}$\\
Arghadip Koner,$^{1}$
Federico Mellini,$^{1}$ 
Sindhana Pannir-Sivajothi,$^{1}$\\
Yong Rui Poh,$^{1}$
Kai Schwennicke,$^{1}$
Kunyang Sun,$^{1}$
Stephan van den Wildenberg,$^{1}$ \\
Dylan Karzen,$^{2}$
Alec Barron,$^{3}$
Joel Yuen-Zhou$^{1\ast}$\\
\\
\normalsize{$^{1}$Department of Chemistry and Biochemistry, University of California San Diego,}\\
\normalsize{La Jolla, CA 92093, USA}\\
\normalsize{$^{2}$Orange Glen High School, Escondido, CA 92027, USA}\\
\normalsize{$^{3}$Center For Research On Educational Equity, Assessment \& Teaching Excellence,}\\ 
\normalsize{University of California San Diego, La Jolla, CA 92093, USA}\\
\\
\normalsize{$^\ast$To whom correspondence should be addressed; E-mail:  joelyuen@ucsd.edu.}
}

\date{}

\usepackage{bbm}
\usepackage[labelfont=bf,labelsep=period]{caption}

\makeatother

\begin{document}

\baselineskip24pt


\maketitle 


\begin{sciabstract} Topological insulators are insulators in the
bulk but feature chiral energy propagation along the boundary. This
property is topological in nature and therefore robust to disorder.
Originally discovered in electronic materials, topologically protected
boundary transport has since been observed in many other physical
systems. Thus, it is natural to ask whether this phenomenon finds
relevance in a broader context. We choreograph a dance in which a
group of humans, arranged on a square grid, behave as a topological
insulator. The dance features unidirectional flow of movement through
dancers on the lattice edge. This effect persists when people are
removed from the dance floor. Our work extends the applicability of
wave physics to the performance arts. \end{sciabstract}

A topological property of an object is one that is unchanged as the
object undergoes continuous deformation, which includes translation,
rotation, stretching/compression, and bending but excludes puncturing,
tearing, and gluing (together different parts of it). More than just
a theoretical concept, this notion can have real-life applications.
Consider a physical material with a topological property. The latter
is resistant to material imperfections that constitute continuous
deformations (though not necessarily in real space). Due to this robustness,
such materials, which are known as topological materials, have garnered
widespread attention over the past several decades \cite{Hasan2010rev,Qi2011rev,Yan2017rev}. 

To date, the most studied topological material has been the topological
insulator \cite{Hasan2010rev}. A topological insulator is insulating
in the bulk but conducting on the boundary. The earliest known topological
insulators are two-dimensional electronic materials that exhibit the
integer quantum Hall effect \cite{vonKlitzing1980}, in which the
(transverse Hall) conductance along the sample edge is proportional
to a nonzero integer $\nu$. Reflecting the net number of edge states
that support clockwise (or counterclockwise) current, $\nu$ and thus
the edge conductance are topological properties \cite{Thouless1982}.
Remarkably, these characteristics of the edge are intimately related
to properties of the material bulk. Such bulk-boundary correspondence
is a hallmark of topological insulators. 

Since their discovery, topological insulators have been observed in
a plethora of other physical media. Examples include traditional wave
media, both natural (e.g., oceanic and atmospheric fluids \cite{Delplace2017})
and synthetic (e.g., photonic \cite{Wang2009,Lu2014rev} and acoustic
\cite{He2016,Ma2019rev} lattices). Topological insulators have also
been reported in settings that have less in common with electronic
materials: systems governed by Newton's equations of motion \cite{Susstrunk2015,Nash2015,Ma2019rev},
amorphous materials \cite{Mitchell2018}, active matter \cite{Shankar2017,Dasbiswas2018,Shankar2022rev},
and stochastic processes \cite{Murugan2017,Dasbiswas2018,Tang2021}. 

The ubiquity of topological insulators prompts the question of whether
their physics can manifest in contexts that transcend the usual boundaries
of science. In this work, we present a human topological insulator
in the form of a group dance. Functioning literally as a numerical
integrator of the time-dependent Schr\"odinger equation (TDSE), the
dance features chiral motion through people along the edge of the
dance floor, even when ``defects'' are introduced by removing dancers.
In essence, this dance is distinct from those that serve as natural
examples or purely qualitative representations of concepts in science
and math (including seismic waves \cite{Miller2012}, electrical circuits
\cite{NOVAPBS2016}, and topology \cite{Graf2013}). Thus, the dance
described in this article serves both as a rigorous realization of
topological edge modes as well as an ideal outreach activity to introduce
broader audiences to the universal concepts of topological protection.

\begin{figure}[p]
\centering \includegraphics[width=4.75in]{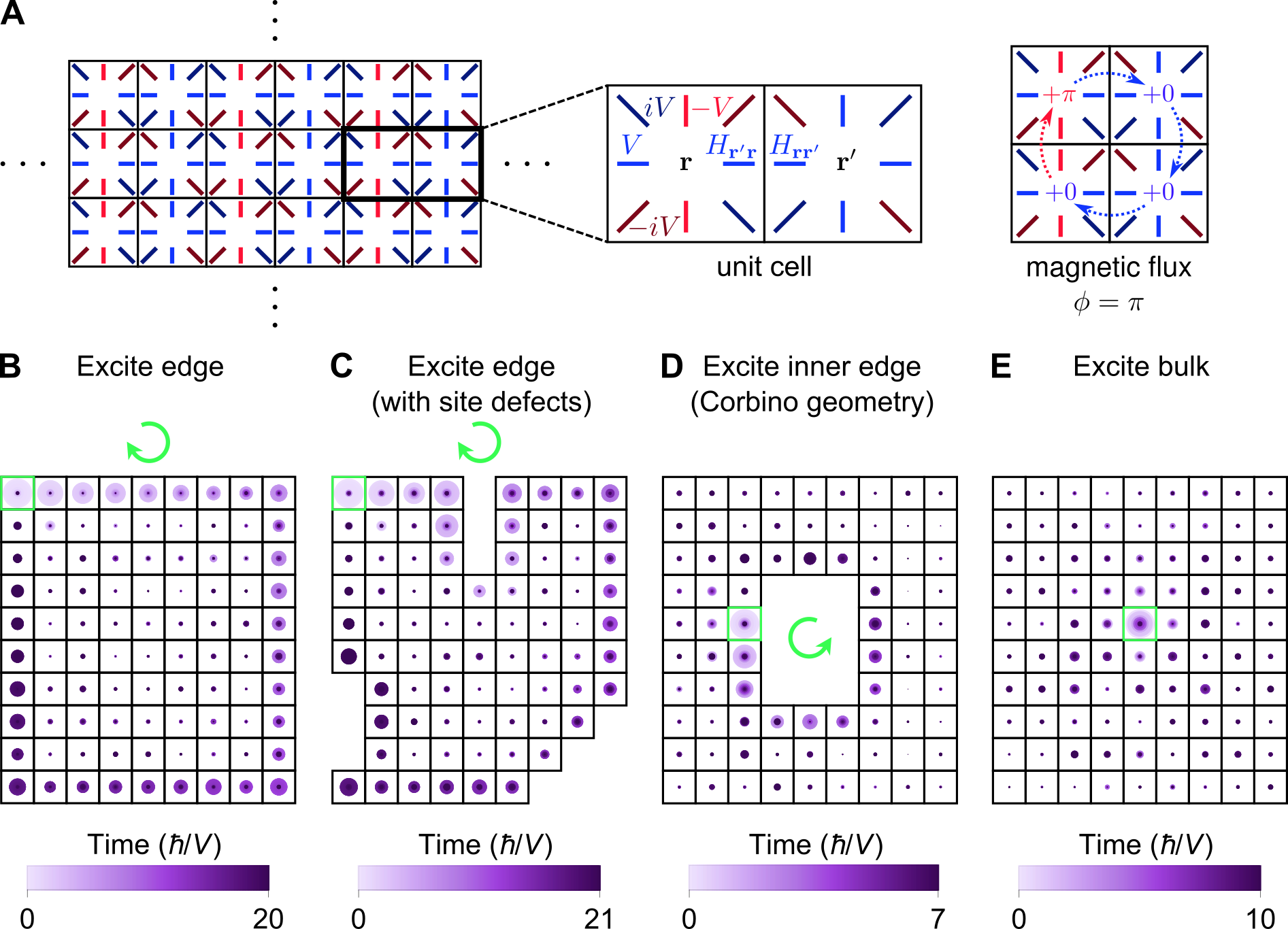}\caption{\textbf{Dynamics of a model topological insulator.} (\textbf{A}) Pictorial
representation of the Harper-Hofstadter Hamiltonian ($H$) with next-nearest-neighbor
hopping and magnetic flux $\phi=\pi$ (Eq. \ref{eq:h}). (\textbf{B}
to \textbf{E}) Dynamics of $H$ on a 10 $\times$ 9 lattice. The system
is excited at a site (green box) located (B) on the edge, (C) on the
edge of the lattice with site defects, (D) on the inner edge of the
lattice with a 4 $\times$ 3 hole in the middle (i.e., Corbino geometry),
and (E) in the bulk. Site probabilities at different times are overlaid
in chronological order (i.e., later times on top). The probability
of the system being at each site is represented by a circle (area
$\propto$ probability). Excitations move unidirectionally along each
edge, where the chirality of motion is indicated by a green arrow.\label{fig:dynamics}}
\end{figure}

To begin the choreography, we consider the Harper-Hofstadter Hamiltonian
\cite{Harper1955,Hofstadter1976} with next-nearest neighbor (NNN)
coupling \cite{Hatsugai1990} and magnetic flux $\phi=\pi$ per plaquette
(Fig. \ref{fig:dynamics}A),
\begin{align}
H & =V\sum_{m,n}\biggl[\left(|m+1,n\rangle\langle m,n|+e^{i\phi m}|m,n+1\rangle\langle m,n|\right.\nonumber \\
 & \quad\left.+e^{i\phi(m+1/2)}|m+1,n+1\rangle\langle m,n|\right.\nonumber \\
 & \quad\left.+e^{i\phi(m-1/2)}|m-1,n+1\rangle\langle m,n|\right)+\text{H.c.}\biggr],\label{eq:h}
\end{align}
which models an electron hopping on a square lattice in a magnetic
field. The lattice sites are labeled by $\mathbf{r}=(m,n)$, and $V$($>0$
here) is the magnitude of intersite coupling. Hops to a nearest neighbor
(NN) occur with an amplitude of $\pm V$, while hops to a NNN occur
with an amplitude of $\pm iV$. Here, H.c. stands for Hermitian conjugate.

The Hamiltonian $H$ gives rise to several dynamical features that
are characteristic of topological insulators, as shown by simulations
on a finite lattice \cite{MaterialsAndMethods}. When exciting an
edge site of a square-shaped lattice, the excitation propagates clockwise
along the edge (Fig. \ref{fig:dynamics}B and Movie S1). This chiral
transport persists after introducing lattice defects of various shapes
(Fig. \ref{fig:dynamics}C and Movie S2). For a lattice with a hole
in the middle, which is known as the Corbino geometry \cite{Corbino1911,Halperin1982},
an excitation at the inner edge moves along this edge with opposite
handedness, i.e., counterclockwise (Fig. \ref{fig:dynamics}D and
Movie S3). The unidirectional conduction on the edges is drastically
different from the dynamics in the bulk, in which a localized excitation
diffuses with little directional selectivity (Fig. \ref{fig:dynamics}E
and Movie S4). 

\begin{figure}
\centering\includegraphics[width=4.75in]{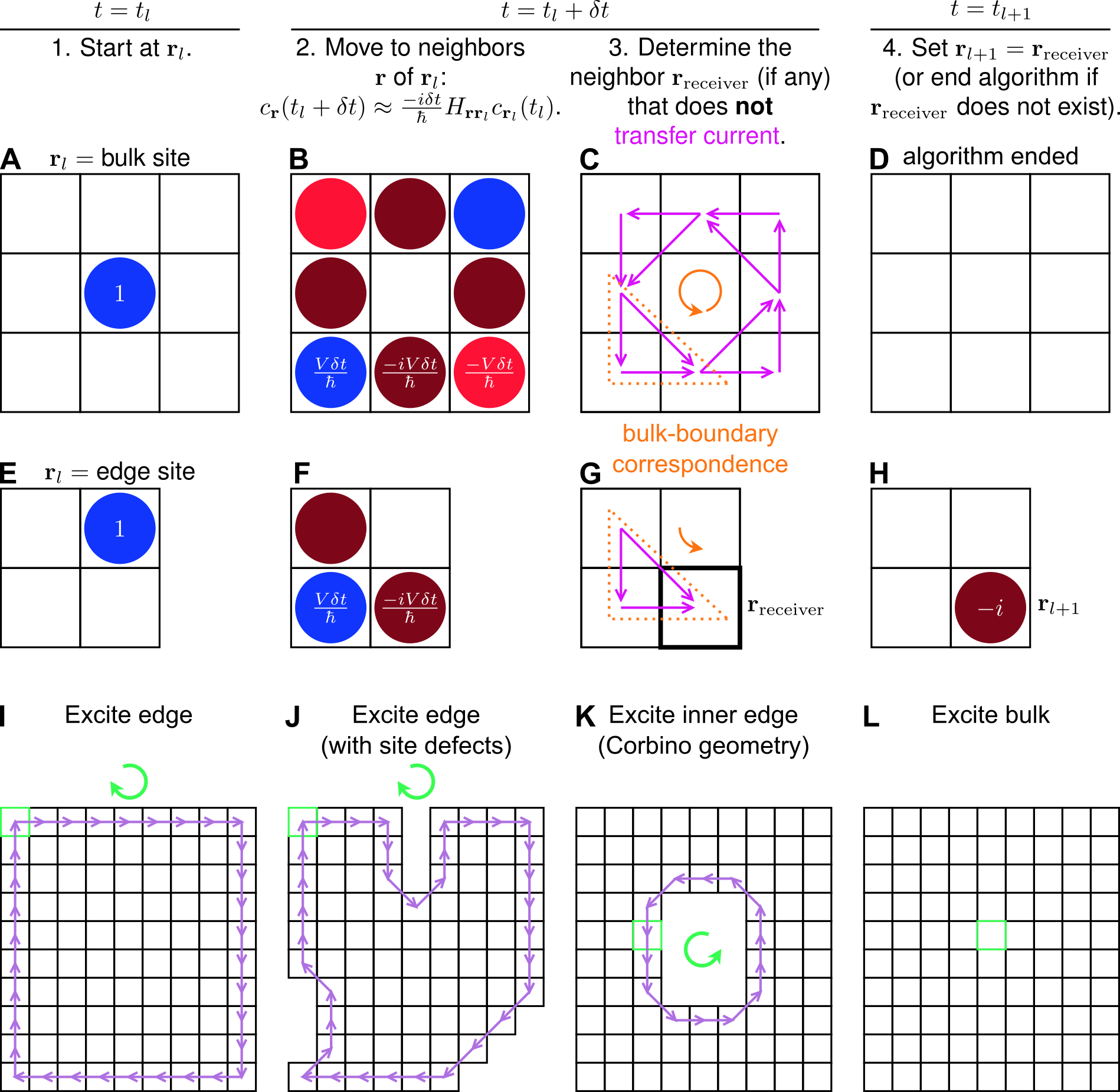}

\caption{\textbf{Algorithm to generate discrete-time dynamics of a topological
insulator.} (\textbf{A} to \textbf{H}) Illustration of the algorithm.
The wavefunction starts an iteration either (A to D) in the bulk or
(E to H) on the edge. (C and G) For each pair of sites $\mathbf{r}$
and $\mathbf{r}'$ such that $\mathbf{r}$ transfers current to $\mathbf{r}'$,
the current vector $\langle J_{\mathbf{r}\rightarrow\mathbf{r}'}(t_{l}+\delta t)\rangle(\mathbf{r}'-\mathbf{r})$
is represented by a purple arrow. There is a bulk-boundary correspondence
with respect to the current field (orange triangles) and its chirality
(orange arrows). (\textbf{I} to \textbf{L}) Dynamics simulated by
the algorithm, where the excitation conditions and lattice geometries
are those of Fig. \ref{fig:dynamics}\textit{, }B to E, respectively.
For each simulation, the wavefunction starts at the site indicated
by the green box. A purple arrow represents the movement of the wavefunction
from $\mathbf{r}_{l}$ (tail) at time step $l$ to $\mathbf{r}_{l+1}$
(head) at time step $l+1$. The discrete-time dynamics shows unidirectional
motion along each edge, where the chirality of motion is indicated
by a green arrow.\label{fig:algorithm}}
\end{figure}

To capture such dynamics in a dance, we first present an algorithm
to (approximately) propagate the wavefunction $|\psi(t)\rangle=\sum_{\mathbf{r}}c_{\mathbf{r}}(t)|\mathbf{r}\rangle$
in discrete time. The algorithm goes as follows (Fig. \ref{fig:algorithm},
A to H):
\begin{enumerate}
\item At the $l$th time step, $t=t_{l}$, the wavefunction is at site $\mathbf{r}_{l}$:
\begin{equation}
|\psi(t_{l})\rangle=c_{\mathbf{r}_{l}}(t_{l})|\mathbf{r}_{l}\rangle,\label{eq:psi_t_l}
\end{equation}
where
\begin{equation}
c_{\mathbf{r}}(t_{l})=\begin{cases}
\pm1, & \sigma(\mathbf{r}_{l})\text{ even},\\
\pm i, & \sigma(\mathbf{r}_{l})\text{ odd},
\end{cases}\label{eq:c_rl}
\end{equation}
and $\sigma(\mathbf{r})=m+n$ (Fig. \ref{fig:algorithm}, A and E).
\item Evolve the wavefunction forward by time $\delta t<t_{l+1}-t_{l}$,
and approximate the resulting state up to $O(\delta t)$: 
\begin{align}
|\psi(t_{l}+\delta t)\rangle & \approx\left(1-\frac{i\delta t}{\hbar}H\right)|\psi(t_{l})\rangle\nonumber \\
 & =c_{\mathbf{r}_{l}}(t_{l})\left(|\mathbf{r}_{l}\rangle-\frac{i\delta t}{\hbar}\sum_{\mathbf{r}\in\mathcal{N}(\mathbf{r}_{l})}H_{\mathbf{r}\mathbf{r}_{l}}|\mathbf{r}\rangle\right),\label{eq:psi_t-p-dt}
\end{align}
where $\mathcal{N}(\mathbf{r}_{l})$ is the set of neighbors (NN and
NNN) of $\mathbf{r}_{l}$ (Fig. \ref{fig:algorithm}, B and F).
\item Determine the neighbor $\mathbf{r}_{\text{receiver}}$ (if any) of
$\mathbf{r}_{l}$ that does not transfer current to any site (Fig.
\ref{fig:algorithm}, C and G). Here, the (probability) current from
$\mathbf{r}$ to $\mathbf{r}'$ is represented by the operator $J_{\mathbf{r}\rightarrow\mathbf{r}'}=\frac{i}{\hbar}\left(H_{\mathbf{r}\mathbf{r}'}|\mathbf{r}\rangle\langle\mathbf{r}'|-H_{\mathbf{r}'\mathbf{r}}|\mathbf{r}'\rangle\langle\mathbf{r}|\right)$
\cite{Baranger1991,Todorov2002}. We say that $\mathbf{r}$ transfers
current to $\mathbf{r}'$ if $\langle J_{\mathbf{r}\rightarrow\mathbf{r}'}(t_{l}+\delta t)\rangle>0$.
\item If there is a neighbor $\mathbf{r}_{\text{receiver}}$ of $\mathbf{r}_{l}$,
set
\begin{equation}
|\psi(t_{l+1})\rangle=\text{sgn}\left[c_{\mathbf{r}_{\text{receiver}}}(t_{l}+\delta t)\right]|\mathbf{r}_{\text{receiver}}\rangle,
\end{equation}
where $\text{sgn}\,z=z/|z|$ is the complex sign function, and $\mathbf{r}_{l+1}=\mathbf{r}_{\text{receiver}}$
(Fig. \ref{fig:algorithm}H); return to Step 2. If not, the algorithm
terminates (Fig. \ref{fig:algorithm}D).
\end{enumerate}
Crucial to the algorithm are Steps 3 and 4, during which the probability
amplitudes interfere, ultimately localizing at $\mathbf{r}_{\text{receiver}}$.
This makes intuitive sense, since $\mathbf{r}_{\text{receiver}}$
is the ``attractor/sink'' of the current field at time $t_{l}+\delta t$
(Step 3, Fig. \ref{fig:algorithm}G). We have assumed that there is
at most one neighbor $\mathbf{r}_{\text{receiver}}$ of $\mathbf{r}_{l}$,
which is true for the lattice geometries appearing in this work {[}see
supplementary materials (SM) Section \ref{sec:multiple-rns}{]}. 

In Fig. \ref{fig:algorithm}, I to L, we show the dynamics generated
by the algorithm. The results are in excellent qualitative agreement
with the exact dynamics (Fig. \ref{fig:dynamics}, B to E, respectively;
see also Movies S1-S4, respectively). Notably, the algorithm reproduces
the confinement of an edge excitation to the edge, the chirality with
which this excitation moves, and the robustness of these properties
to site defects. Also captured is the diagonal movement of an edge
excitation as it travels around the defects (cf. Figs. \ref{fig:algorithm}J
and \ref{fig:dynamics}C) and, in the Corbino geometry, past the corners
of the inner edge (cf. Figs. \ref{fig:algorithm}K and \ref{fig:dynamics}D). 

Underlying the high qualitative accuracy of Algorithm 1 is the direct
dependence of the dynamics on the site currents. For an iteration
starting at a bulk site (Fig. \ref{fig:algorithm}A), current flows
to and from all neighbors of the initial site (Fig. \ref{fig:algorithm}C).
As a result, the algorithm ends (Fig. \ref{fig:algorithm}, D and
L), reflecting the absence of unidirectional propagation in the bulk.
Interestingly though, the current vectors form a vortex with a well-defined
chirality (i.e., counterclockwise; see Fig. \ref{fig:algorithm}C,
purple arrows). By considering an appropriate subset of this current
field (Fig. \ref{fig:algorithm}C, orange dotted triangle), one can
obtain the current field for an iteration beginning at an edge (Fig.
\ref{fig:algorithm}G, orange dotted triangle). This subfield (Fig.
\ref{fig:algorithm}G, purple arrows) determines the site that the
wavefunction will occupy at the start of the next iteration (Fig.
\ref{fig:algorithm}H). Thus, the currents of a bulk-localized wavefunction
indicate how an edge-localized wavefunction would propagate. In particular,
the subset relation between edge and bulk current fields (Fig. \ref{fig:algorithm},
G and C, orange dotted triangles), and the structure of the latter
(Fig. \ref{fig:algorithm}C, purple arrows), explain why edge excitations
are confined to the edge. Furthermore, the chirality of the bulk currents
(Fig. \ref{fig:algorithm}C, orange arrow) is directly correlated
with the chirality of the edge dynamics (Fig. \ref{fig:algorithm}G,
orange arrow). These properties constitute a dynamical form of bulk-boundary
correspondence. In particular, they closely resemble the classical
picture of an electron in a magnetic field, where the circular orbits
of a free particle manifest as unidirectional skipping motion along
an edge \cite{Hasan2010rev}. 

To convert the algorithm to a dance, it is useful to transform the
wavefunction from the complex plane to the real numbers. To see that
this transformation is possible, notice that, at all times $t$ explicitly
considered in the algorithm (i.e., $t_{l},t_{l}+\delta t$), the probability
amplitude at each $\mathbf{r}$ satisfies
\begin{equation}
c_{\mathbf{r}}\in\begin{cases}
\mathbbm{R}, & \sigma(\mathbf{r})\text{ even},\\
i\mathbbm{R}, & \sigma(\mathbf{r})\text{ odd}.
\end{cases}\label{eq:c_r-sign}
\end{equation}
This property results from the choice of initial state (Eq. \ref{eq:c_rl}),
which depends on whether $\sigma(\mathbf{r})$ is even or odd; the
update rule of Step 4 (Eq. \ref{eq:psi_t-p-dt}); and the structure
of the Hamiltonian (Eq. \ref{eq:h}), which has purely real NN couplings
and purely imaginary NNN couplings. Moreover, since all hopping amplitudes
have the same magnitude (Eq. \ref{eq:h}), $|c_{\mathbf{r}}(t_{l}+\delta t)|=|c_{\mathbf{r}'}(t_{l}+\delta t)|$
for all neighboring sites $\mathbf{r},\mathbf{r}'$of $\mathbf{r}_{l}$.
It follows that only the signs of these coefficients are necessary
to capture the essential physics, where the signs are given by
\begin{equation}
f(z)=\begin{cases}
\text{sgn}(z), & z\in\mathbbm{R},\\
\text{sgn}(z/i), & z\in i\mathbbm{R}.
\end{cases}\label{eq:f}
\end{equation}
Thus, applying $f$ to all probability amplitudes $c_{\mathbf{r}}$
recasts the algorithm in terms of real numbers (SM Section \ref{sec:algorithm2}),
i.e., the transformed amplitudes $c_{\mathbf{r}}'\equiv f(c_{\mathbf{r}})$
and ``effective Hamiltonian''
\begin{equation}
\mathcal{H}_{\mathbf{r}'\mathbf{r}}=\begin{cases}
-f(H_{\mathbf{r}'\mathbf{r}}), & \sigma(\mathbf{r}')\text{ odd and }\sigma(\mathbf{r})\text{ even},\\
f(H_{\mathbf{r}'\mathbf{r}}), & \text{else}.
\end{cases}\label{eq:h_eff}
\end{equation}
By definition, $c_{\mathbf{r}}'$ and $\mathcal{H}_{\mathbf{r}'\mathbf{r}}$
each takes the values $0,\pm1$. We emphasize that the site probabilities
at times $t_{l}$, and hence the overall dynamics, remain unchanged
from the original algorithm. 

\begin{figure}
\centering\includegraphics[width=4.75in]{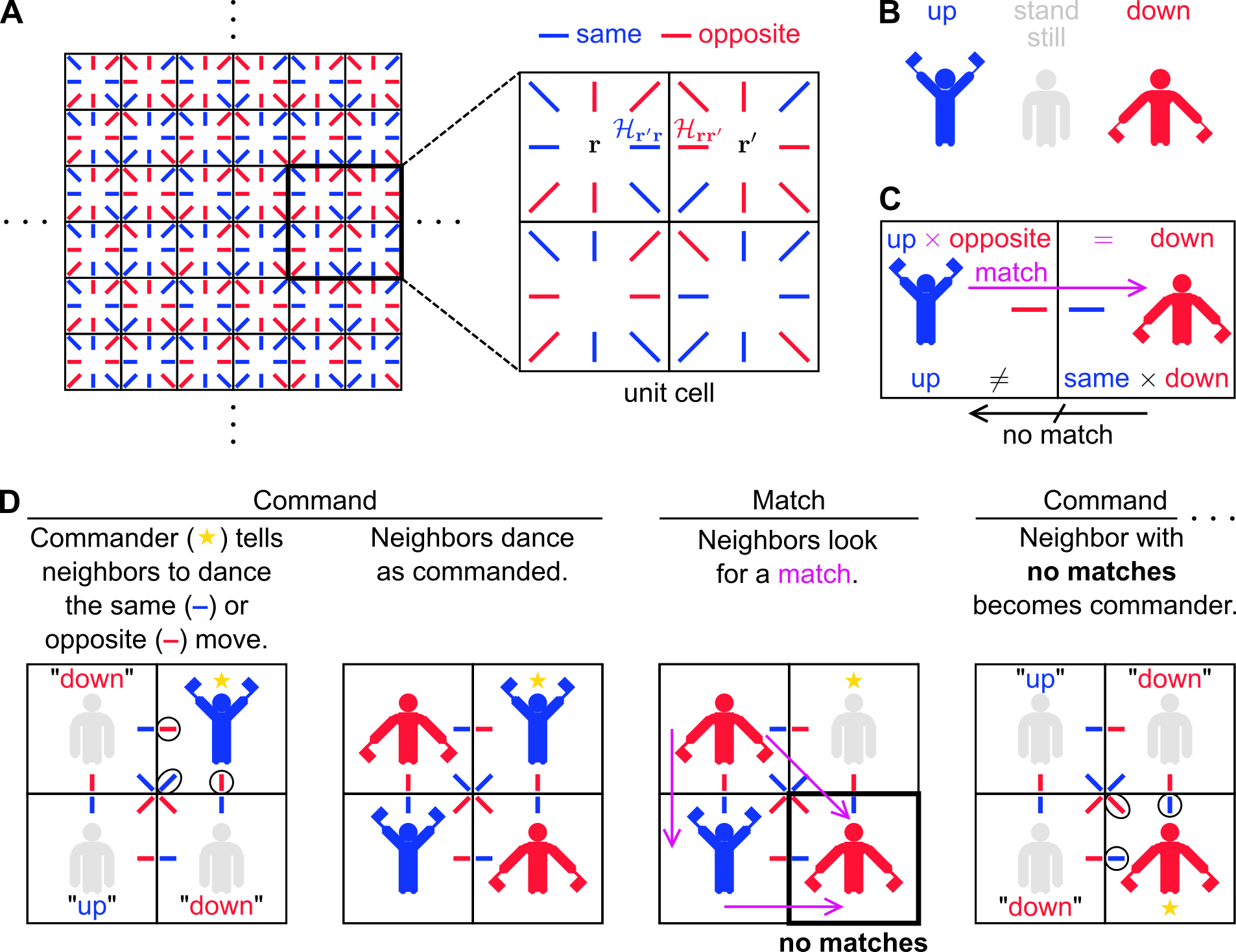}

\caption{\textbf{Mechanics of the dance.} (\textbf{A}) Dance floor. As shown
in the zoom-in of the unit cell, the squares represent the lattice
sites and the colored lines represent the matrix elements of $\mathcal{H}$
(Eq. \ref{eq:h_eff}), the ``effective Hamiltonian'' that generates
the dance dynamics. (\textbf{B}) Dance moves. (\textbf{C}) Example
of what a ``match'' is and is not. \textit{(}\textbf{D}) Illustration
of the dance steps for one round of the dance. \label{fig:dance-mechanics}}
\end{figure}

We proceed to choreograph a dance via a direct mapping of the reformulated
algorithm. The probability amplitudes are represented by dance moves:
\begin{equation}
c_{\mathbf{r}}'=\begin{cases}
\quad1 & \rightarrow\quad\text{up},\\
\quad0 & \rightarrow\quad\text{stand still,}\\
-1 & \rightarrow\quad\text{down}.
\end{cases}\label{eq:dance-moves}
\end{equation}
``Up'' and ``down'' refer to the waving of flags with arms pointed
in the indicated direction (Fig. \ref{fig:dance-mechanics}B, blue
and red, respectively). In contrast, ``stand still'' is exactly
as the name suggests, with arms relaxed at the sides of the dancer
(Fig. \ref{fig:dance-mechanics}B, gray). The (nonzero) hopping amplitudes
are represented as
\begin{equation}
\mathcal{H}_{\mathbf{r}'\mathbf{r}}=\begin{cases}
\quad1 & \rightarrow\quad\text{same},\\
-1 & \rightarrow\quad\text{opposite}.
\end{cases}\label{eq:same-opposite}
\end{equation}
The above redefinitions result in the following rules for multiplying
the probability amplitudes with the hopping amplitudes:
\begin{equation}
c_{\mathbf{r}}'\times\mathcal{H}_{\mathbf{r}'\mathbf{r}}=\begin{cases}
\text{up}\times\text{same} & =\quad\text{up},\\
\text{up}\times\text{opposite} & =\quad\text{down},\\
\text{down}\times\text{same} & =\quad\text{down},\\
\text{down}\times\text{opposite} & =\quad\text{up}.
\end{cases}\label{eq:multiply}
\end{equation}

With this ``human representation'' of the wavefunction and ``effective
Hamiltonian'' $\mathcal{H}$, the real-valued algorithm is readily
converted to a dance, described as follows. The dance floor (Fig.
\ref{fig:dance-mechanics}A) is a finite square grid, where the squares
contain blue and red lines. Each square represents a lattice site
$\mathbf{r}$. The blue/red lines within that square represent the
amplitudes $\mathcal{H}_{\mathbf{r}'\mathbf{r}}$ = same/opposite
of hopping to neighboring sites $\mathbf{r}'$ (Fig. \ref{fig:dance-mechanics}A,
zoom-in). For each site that the electron can occupy, a dancer is
placed in the corresponding square. Given this setup, the dance is
performed in rounds, where each round has the steps below (Fig. \ref{fig:dance-mechanics}D):
\begin{lyxlist}{00.00.0000}
\item [{(Command)}] The person designated as the \textit{commander} is
dancing up or down. The commander tells each neighbor to dance in
the same or opposite way, according to the line in the commander's
square that points to this neighbor (Fig. \ref{fig:dance-mechanics}D,
leftmost panel, circled lines). The neighbors start dancing as commanded. 
\item [{(Command-to-Match\ transition)}] The commander stands still. 
\item [{(Match)}] Within the neighbors of the commander, each person scans
across the others, looking for a \textit{match}. As demonstrated in
Fig. \ref{fig:dance-mechanics}C, the person at $\mathbf{r}$ \textit{matches}
with the person at $\mathbf{r}'$ if the dance move of the former,
times $\mathcal{H}_{\mathbf{r}'\mathbf{r}}$, equals the dance move
of the latter, where $\mathcal{H}_{\mathbf{r}\mathbf{r}_{\text{c}}}$
is given by the line in square $\mathbf{r}$ that points to square
$\mathbf{r}'$.
\item [{(Match-to-Command\ transition)}] All people with a match stop
dancing. If there is a person without a match, this person continues
dancing and becomes the commander; return to the Command step. If
everyone has a match, the dance ends. 
\end{lyxlist}
The Command step (equivalent to Step 2 of the algorithm) represents
the spreading of a localized wavefunction to neighboring sites. The
probability amplitudes at these sites interfere during the Match step
(equivalent to Step 3 of the algorithm) and Match-to-Command transition
(equivalent to Step 4 of the algorithm). Rather than having to memorize
the values of $\mathcal{H}_{\mathbf{r}'\mathbf{r}}$ for Command and
Match, the dancers simply consult the blue and red lines in their
respective squares.

\begin{figure}
\centering\includegraphics[width=4.75in]{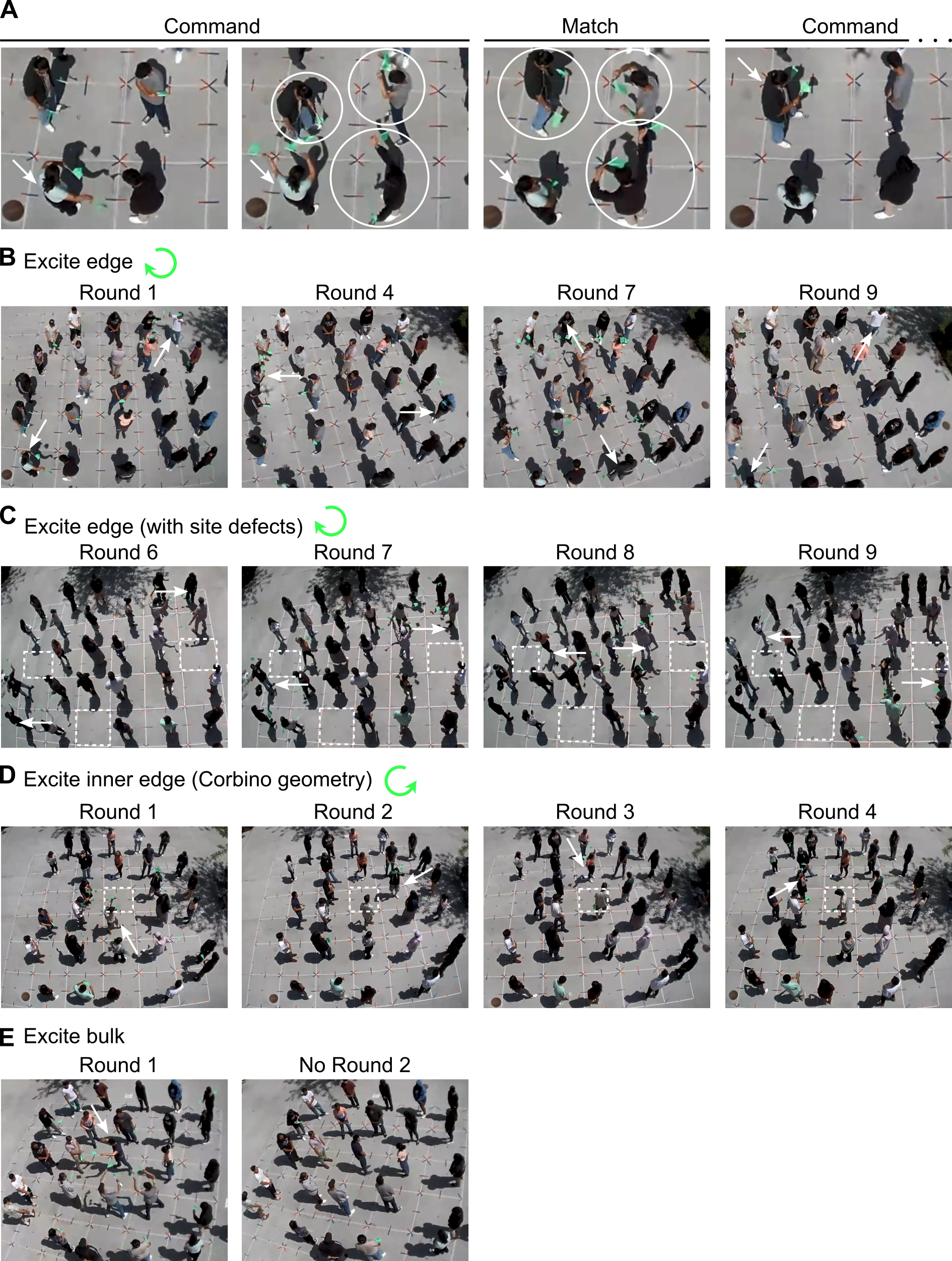}

\caption{\textbf{Dynamics of a dance-based human topological insulator.} (\textbf{A})
Snapshots from one round of the dance. (\textbf{B} to \textbf{E})
Snapshots of the dance where the initial (Round 1) dancers (i.e.,
commanders) are (B) on the lattice edge, (C) on the edge of a lattice
with site defects, (D) on the inner edge of a lattice with a hole
in the middle (i.e., Corbino geometry), and (E) in the lattice bulk.
For dances where the initial dancers are on the edge, the green arrow
indicates the chirality with which the dancing propagates. White dashed
boxes indicate sites without a dancer. In (A to E), white arrows indicate
commanders. In (A), the white circles indicate neighbors (NN and NNN)
of the commander who are dancing up or down.\label{fig:dance-dynamics}}
\end{figure}

As a science outreach event, we taught the dance to students at Orange
Glen High School in Escondido, California \cite{MaterialsAndMethods}.
Overall, the students mastered the dance steps (Fig. \ref{fig:dance-dynamics}A)
in under one hour. The students (plus some of us) then performed the
dance for various initial conditions and lattice geometries \cite{MaterialsAndMethods}.
To engage more students, some performances began with two people dancing
(as commander) on the same dance floor; two independent dances ensued
simultaneously (see, for example, Fig. \ref{fig:dance-dynamics}B
and Movie S5) for all but one dance round (see below). The performances
display key dynamical features of topological insulators, namely,
those generated by the algorithm on which the choreography is based.
When the initial dancers are at an edge, the dancing propagates unidirectionally
along this edge: clockwise on the outer edge of a square-shaped lattice
(Fig. \ref{fig:dance-dynamics}B and Movie S5) and counterclockwise
on the inner edge of a lattice with Corbino geometry (Fig. \ref{fig:dance-dynamics}D
and Movie S7). For the former lattice, we introduced site defects
by removing people at the edge of the dance floor. Still, the lattice
sustains an edge-confined and clockwise-oriented ``dance wave,''
which maneuvers around the vacancies (Fig. \ref{fig:dance-dynamics}C
and Movie S6) and even persists through the interference of two concurrent
dances (Movie S6) . As expected (Fig. \ref{fig:algorithm}L), the
dance only lasts one round when an initial dancer is in the bulk (Fig.
\ref{fig:dance-dynamics}E and Movie S8). 

In summary, we have choreographed a dance in which a group of people
behave as a topological insulator. The choreography involves developing
an algorithm for approximate wavefunction propagation and mapping
the wavefunction first to the real numbers and then to human movements.
The resulting dance, which operates as a numerical integrator of the
TDSE, exhibits the salient dynamics of topological insulators. This
work provides a blueprint for creating a classical simulator of topological
insulators. Achieving this task for additional Hamiltonians would
mark an intriguing and unique frontier at the interface of wave physics,
science education, and performance arts. 

\bibliographystyle{Science}

\input{topological-dance_v2.bbl}
\section*{Acknowledgments}

M.D. would like to thank Raphael Ribeiro for discussions on topological
insulators at the beginning of this project. We are grateful to the
UCSD Chem 126a Fall 2019 undergraduate students for participating
in the trial lesson for a preliminary version of the dance, and to
Luis Mart\'inez-Mart\'inez and Leonardo Calder\'on for helping
run that trial lesson. We thank Madison Edwards, Hilliary Frank, Melanie
Gonzalez, Etienne Palos, Richa Rashmi, Michael Reiss, Shubham Sinha,
Luisa Andreuccioli, Benjamin Huang, and \textcolor{black}{Clara van
den Wildenberg}\textcolor{red}{{} }for pariticipating in the practice
lessons for the final version of the dance. Regarding the official
dance performances, we are grateful to Dania Monroy and the students
of D.K. for their participation. We thank\textcolor{black}{{} Amy Booth,
Shannon Chamberlin, and Susan Yonezawa for their help in organizing
the outreach event.}

\subsection*{Funding}

The scientific and outreach components of this work were funded by
the NSF Grant No. CAREER CHE 1654732. 

\subsection*{Author contributions}

J.Y.-Z. conceptualized and supervised the study. M.D. designed the
wavefunction propagation algorithms and ran simulations. M.D. choreographed
the dance, with input from J.B.P.-S., J.A.C.-G.-A., A.K., F.M., S.P.-S.,
Y.R.P., K. Schwennicke, K. Sun, S.v.d.W., and J.Y.-Z. M.D., D.K.,
A.B., and J.Y.-Z. organized the outreach event. M.D., J.B.P.-S., J.A.C.-G.-A.,
A.K., F.M., S.P.-S., Y.R.P., K. Schwennicke, K. Sun, S.v.d.W., J.Y.-Z.,
and D.K. ran the event. J.B.P.-S. and S.v.d.W. took pictures and recorded
videos of the dance lessons and performances. M.D. and J.B.P.-S. analyzed
the dance performances. M.D. and J.Y.-Z. wrote the manuscript.

\subsection*{Competing interests}

The authors declare no competing interests.

\subsection*{Data and materials availability}

\textcolor{black}{All data are available in the main text or the supplementary
materials.}

\section*{Supplementary materials}

Materials and Methods\\
Supplementary Text\\
\textcolor{black}{Figs. S1 to S5}\\
\textcolor{black}{Movies S1 to S8}


\clearpage{}


\include{sm_topological-dance_v2}
\end{document}

%% file: sm_topological-dance_v2.tex

\baselineskip24pt


\setcounter{figure}{0} \renewcommand{\thefigure}{S\arabic{figure}}
\setcounter{page}{1} \renewcommand{\thepage}{S\arabic{page}}
\setcounter{equation}{0} \renewcommand{\theequation}{S\arabic{equation}}

\maketitle 

\part*{Supplementary Materials for}

\textbf{This PDF file includes:}

Materials and Methods

Supplementary Text

\textcolor{black}{Figs. S1 to S5}\textit{\textcolor{red}{}}\\
\textit{\textcolor{red}{}}\\
\textbf{\textcolor{black}{Other Supplementary Material for this manuscript
includes the following: }}

\textcolor{black}{Movies S1 to S8}

\clearpage{}

\part*{Materials and Methods}

\section*{\uline{Calculating the dynamics generated by \mbox{$H$}}}

At time $t=0$, the system is excited (i.e., initialized) at site
$\mathbf{r}_{I}$, and the wavefunction is given by
\begin{equation}
|\psi(0)\rangle=|\mathbf{r}_{I}\rangle.
\end{equation}
To calculate the wavefunction at later times, we first move to the
eigenbasis: 
\begin{equation}
|\psi(0)\rangle=\sum_{\alpha}c_{\alpha}(0)|\alpha\rangle,
\end{equation}
where $|\alpha\rangle$ is the eigenstate of $H$ with energy $E_{\alpha}$.
We then calculate the wavefunction at time $t$ as 
\begin{equation}
|\psi(t)\rangle=\sum_{\alpha}c_{\alpha}(0)e^{-iE_{\alpha}t/\hbar}|\alpha\rangle.
\end{equation}
Changing back to the position basis, 
\begin{equation}
|\psi(t)\rangle=\sum_{\mathbf{r}}c_{\mathbf{r}}(t)|\mathbf{r}\rangle,
\end{equation}
we obtain the site probabilities $|c_{\mathbf{r}}(t)|^{2}$. Fig.
\ref{fig:dynamics}, B to E, and Movies S1-S4 plot the site probabilities
as a function of time using a time step of 0.1 $\hbar/V$. 

\section*{\uline{Science outreach: topological dance}}

In this section, we describe the science outreach event at Orange
Glen High School on April 27, 2022, in which we taught the dance to
its students. A dance lesson was held during each of three physics
classes in lieu of the normal class activities. In the lesson, students
learned, practiced, and performed the dance. The lesson took most
of the class period, which was 100 minutes long. Since class sizes
were around 20 or less, we (the school teachers and dance instructors)
joined the students in the dance performances. We recommend that the
dance be carried out with at least 25 people, so that the human lattice
has a well-defined bulk (i.e., people who will never dance up or down
if the initial dancers are at the edge).

\begin{figure}
\centering\includegraphics[scale=0.95]{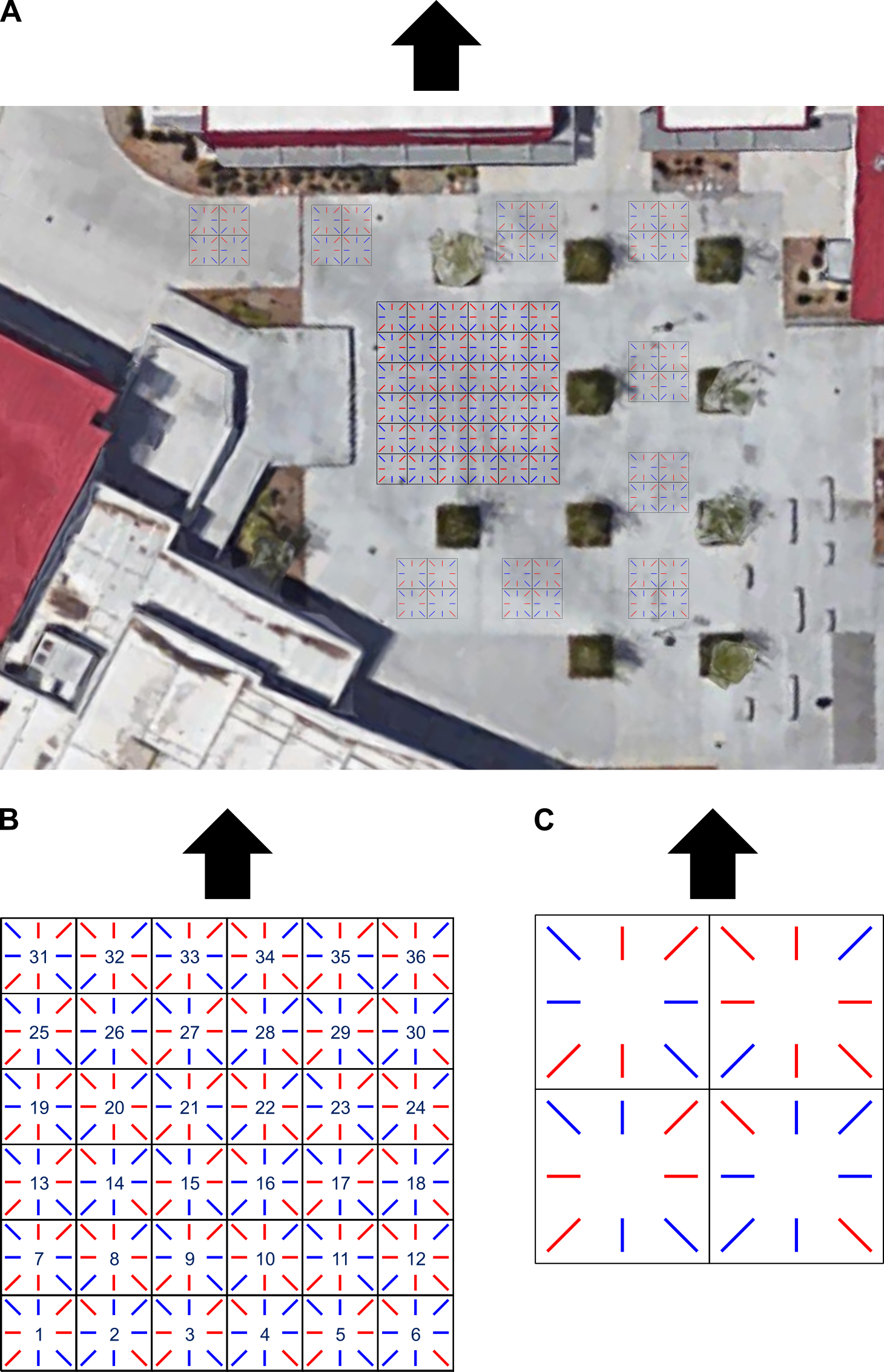}

\caption{\textbf{Sketch of the dance floors.} (\textbf{A} to \textbf{C}) Sketch
of (A) the full setup of 1 big dance floor and 9 small dance floors,
(B) the big dance floor, (C) a small dance floor. The arrows indicate
the orientation of the dance floors in the full setup (A). Note that
the lines are not drawn to scale; see Materials and Methods for the
actual dimensions and Fig. \ref{fig:dance-floors_pic} for the actual
dance floors. \label{fig:dance-floors_sketch}}

\end{figure}

\begin{figure}
\centering\includegraphics{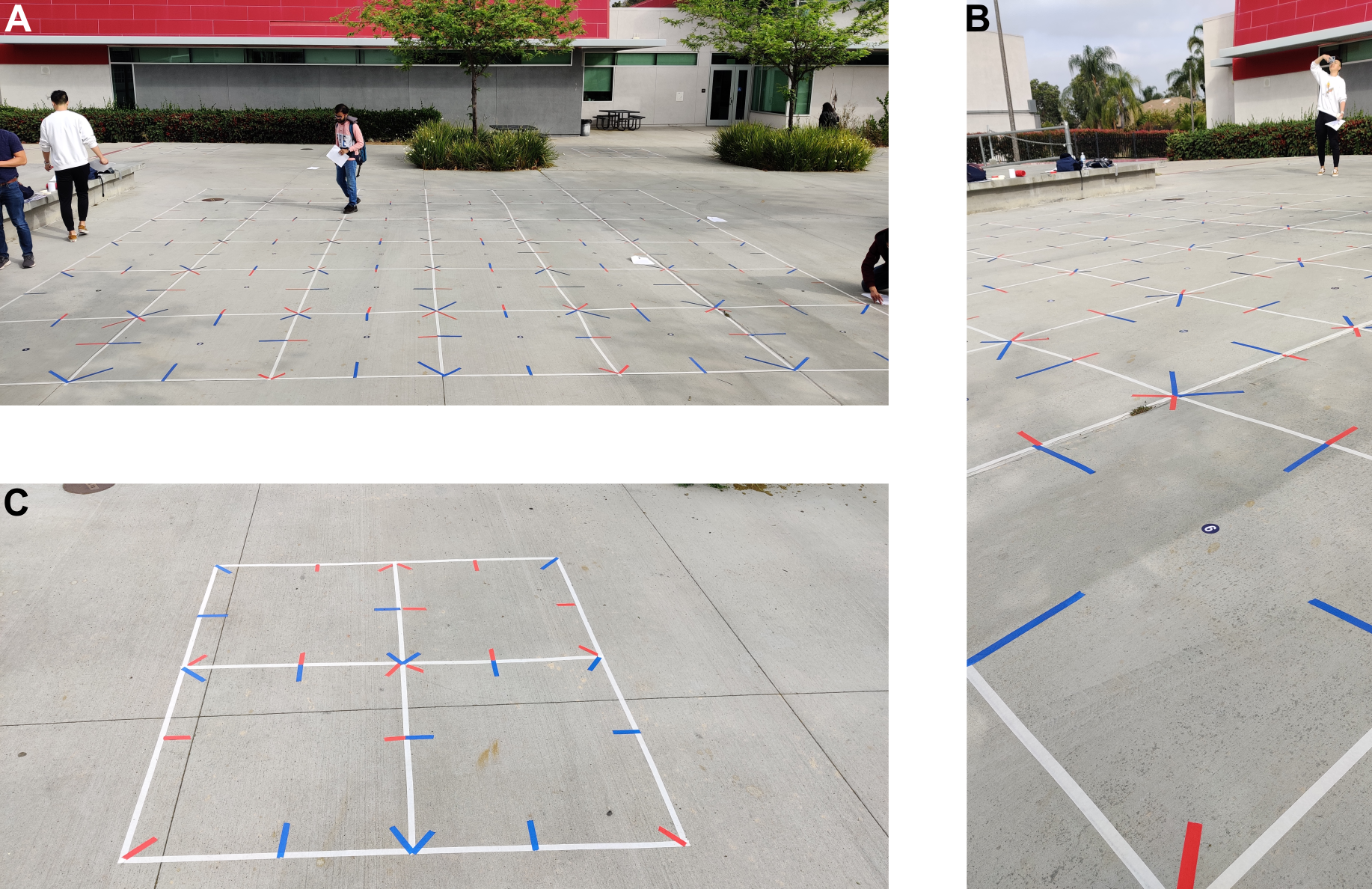}

\caption{\textbf{Pictures of the dance floors.} (\textbf{A} and \textbf{B})
Picture (A) and zoom-in (B) of the big dance floor. (\textbf{C}) Picture
of a small dance floor. \label{fig:dance-floors_pic}}
\end{figure}

Before the lesson, we set up (Fig. \ref{fig:dance-floors_sketch}A)
a big dance floor (6 $\times$ 6 square grid; Figs. \ref{fig:dance-floors_sketch}B
and \ref{fig:dance-floors_pic}A) and several small dance floors (2
$\times$ 2 square grid; Figs. \ref{fig:dance-floors_sketch}C and
\ref{fig:dance-floors_pic}C). The dance floors had grid lines made
of beige masking tape (1'' thick) and squares measuring approximately
1 m $\times$ 1 m (Fig. \ref{fig:dance-floors_pic}). Pieces of blue
and red painters tape (1'' thick and 4-10'' long) were placed in
each square according to the structure of ``effective Hamiltonian''
$\mathcal{H}$ (Fig. \ref{fig:dance-floors_pic}; compare to Fig.
\ref{fig:dance-mechanics}A). In the big dance floor, the squares
were enumerated from 1 to 36 (Fig. \ref{fig:dance-floors_sketch}B
and \ref{fig:dance-floors_pic}B) to assist in the execution of the
dance performances (see below). 

\begin{figure}
\centering\includegraphics{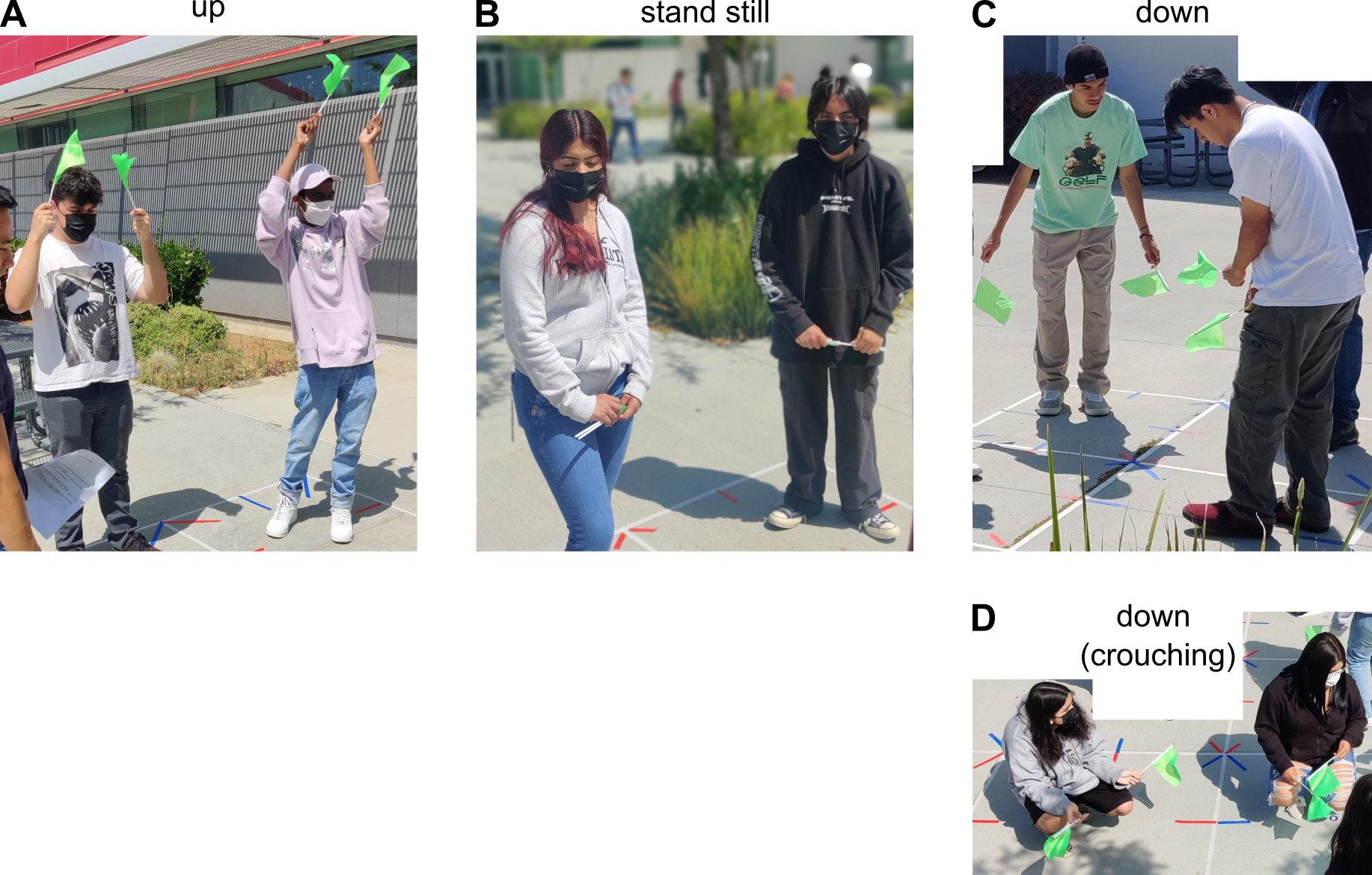}

\caption{\textbf{Pictures of the dance moves.} (\textbf{A}) Up. (\textbf{B})
Stand still. (\textbf{C}) Down (regular version, i.e., while standing).
(\textbf{D}) Down while crouching. \label{fig:dance-moves_pic}}

\end{figure}

Following a brief introduction to topological insulators and the lesson,
the students were divided into groups of 4. Each group moved to a
practice dance floor, where one of us taught them the mechanics of
the dance. The students first learned the dance moves. Fig. \ref{fig:dance-moves_pic}
shows what the dance moves look like in real life (see cartoon version
in Fig. \ref{fig:dance-mechanics}B). To make the moves more distinguishable,
dancers are encouraged to cover their flags with their hands when
doing ``stand still'' (Fig. \ref{fig:dance-moves_pic}B) and crouch
when doing ``down'' (Fig. \ref{fig:dance-moves_pic}D). The students
then learned the Command step. After each student had a chance to
practice being commander, they learned the Match step. Finally, the
students practiced both steps together (including the transitions
between the steps) until mastery was achieved. Most students had mastered
the dance steps after 30-45 minutes. 

Next, the students moved to the main dance floor to rehearse and perform
the dance for 2-3 sets of initial conditions (i.e., who the commanders
are in the first round of the dance) and arrangements of students
(e.g., square-shaped lattice, Corbino geometry). Accompanied by music,
each performance proceeded as follows. We first called out the number(s)
of the student(s) who would serve as the commanders in the first round
of the dance. To begin the dance, we blew a whistle and announced
``Command,'' signaling for the assigned commanders to carry out
Command. Once all neighbors of the commander(s) had begun dancing,
we blew a whistle and announced ``Match,'' initiating the transition
from Command to Match. After 15-20 seconds, which was enough for students
to carry out Match, we blew a whistle and announced ``Command''
to switch to Command of the next round. This cycle was repeated for
each round of the dance.

\part*{Supplementary Text}

\section{A useful property of the current operator}

Consider the operator $J_{\mathbf{r}\rightarrow\mathbf{r}'}$ (see
main text for definition) representing the current from $\mathbf{r}$
to $\mathbf{r}'$. With respect to wavefunction $|\psi(t)\rangle=\sum_{\mathbf{r}}c_{\mathbf{r}}(t)|\mathbf{r}\rangle$,
the expectation value of this current is
\begin{equation}
\langle J_{\mathbf{r}\rightarrow\mathbf{r}'}(t)\rangle=\frac{2}{\hbar}\text{Im}\left[c_{\mathbf{r}'}^{*}(t)H_{\mathbf{r}'\mathbf{r}}c_{\mathbf{r}}(t)\right].\label{eq:j-gen}
\end{equation}
We use this property later in the supplementary text. 

\section{Cases when there are multiple sites $\mathbf{r}_{\text{receiver}}$\label{sec:multiple-rns}}

In the algorithm of the main text, we assume that there is at most
one neighbor (NN or NNN) $\mathbf{r}_{\text{receiver}}$ of $\mathbf{r}_{l}$
that does not transfer current to another neighbor of $\mathbf{r}_{l}$.
The assumption is true for the lattice geometries (e.g., square shaped,
Corbino) employed in our simulations and dances. In this section,
we discuss cases featuring multiple neighbors $\mathbf{r}_{\text{receiver}}$.

\begin{figure}
\centering\includegraphics[width=4.75in]{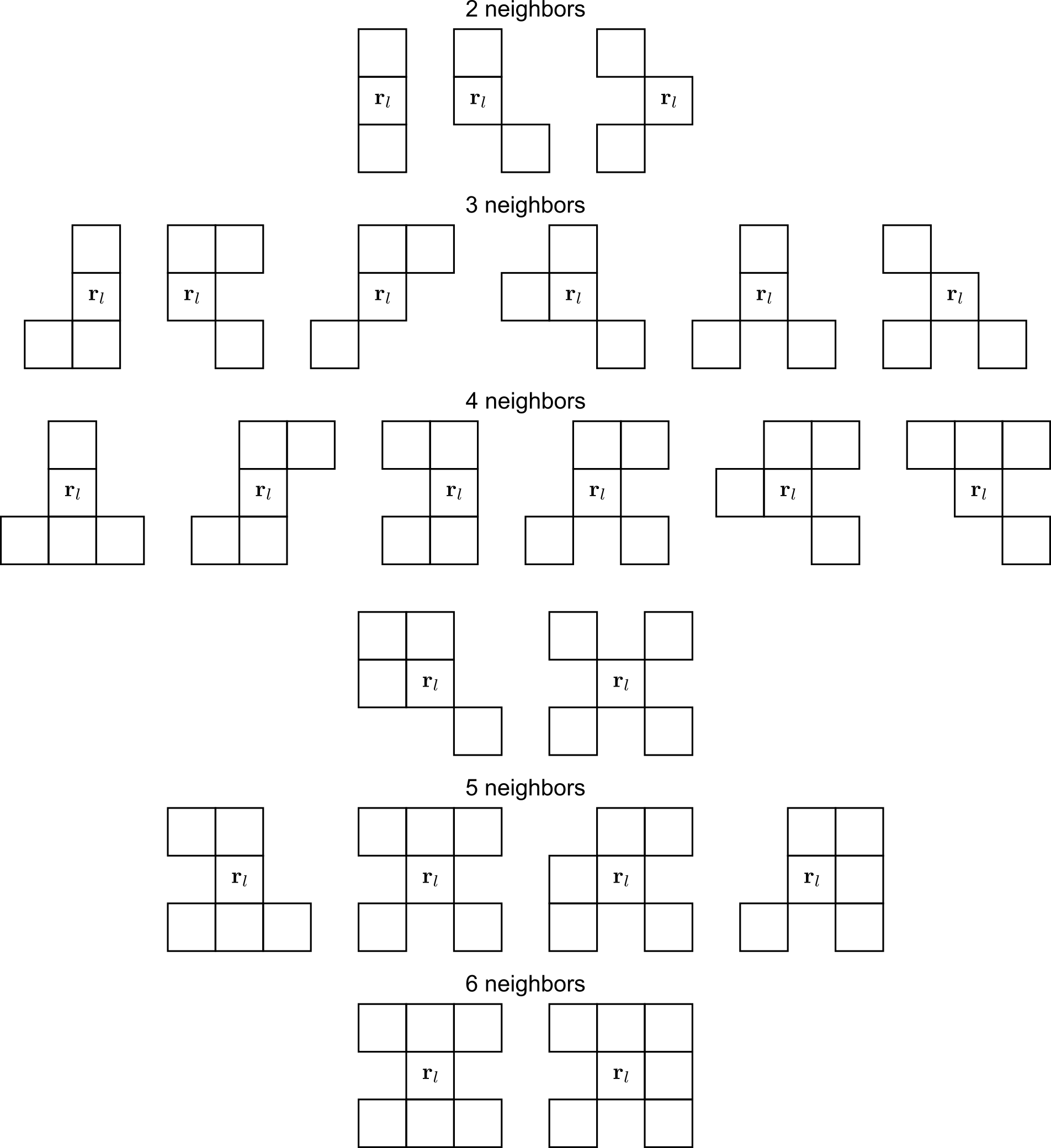}\caption{\textbf{Site arrangements leading to multiple sites }$\mathbf{r}_{\text{receiver}}$\textbf{.}
Arrangements of sites surrounding $\mathbf{r}_{l}$ where the neighbors
of $\mathbf{r}_{l}$ are a union of non-neighboring sets of sites,
leading to multiple sites $\mathbf{r}_{\text{receiver}}$. The arrangements
are listed in increasing order of the number of neighbors surrounding
$\mathbf{r}_{l}$.\label{fig:disjoint_multiple-rns} }

\end{figure}

For a given $\mathbf{r}_{l}$, there are multiple sites $\mathbf{r}_{\text{receiver}}$
when the neighbors of $\mathbf{r}_{l}$ are a union of \textit{non-neighboring}
sets of sites (Fig. \ref{fig:disjoint_multiple-rns}). Here, we denote
sets $S_{1},S_{2},\dots$ of sites as \textit{non-neighboring} if
none of $\mathbf{r}^{(1)},\mathbf{r}^{(2)},\dots$ are neighbors for
any $\mathbf{r}^{(1)}\in S_{1},\mathbf{r}^{(2)}\in S_{2},\dots$.
For a given lattice geometry, the presence of multiple sites $\mathbf{r}_{\text{receiver}}$
can occur if a site and its neighbors form an arrangement of Fig.
\ref{fig:disjoint_multiple-rns}. 

\section{Algorithm 2: generating real-valued, discrete-time dynamics of a
topological insulator \label{sec:algorithm2}}

The algorithm in the main text, hereafter referred to as Algorithm
1, propagates the (complex-valued) wavefunction in discrete time for
the model topological insulator with (complex-valued) Hamiltonian
$H$ (Eq. \ref{eq:h}). In this section, we present the algorithm
that results from transforming the probability amplitudes according
to $c_{\mathbf{r}}\rightarrow c_{\mathbf{r}}'=f(c_{\mathbf{r}})$,
where $f$ is the real-valued function defined in Eq. \ref{eq:f}.
The transformed algorithm, hereafter referred to as Algorithm 2, is
written in terms of real-valued quantities. 

\subsection{Algorithm 2\label{subsec:algorithm2_algorithm}}

\begin{figure}
\centering\includegraphics{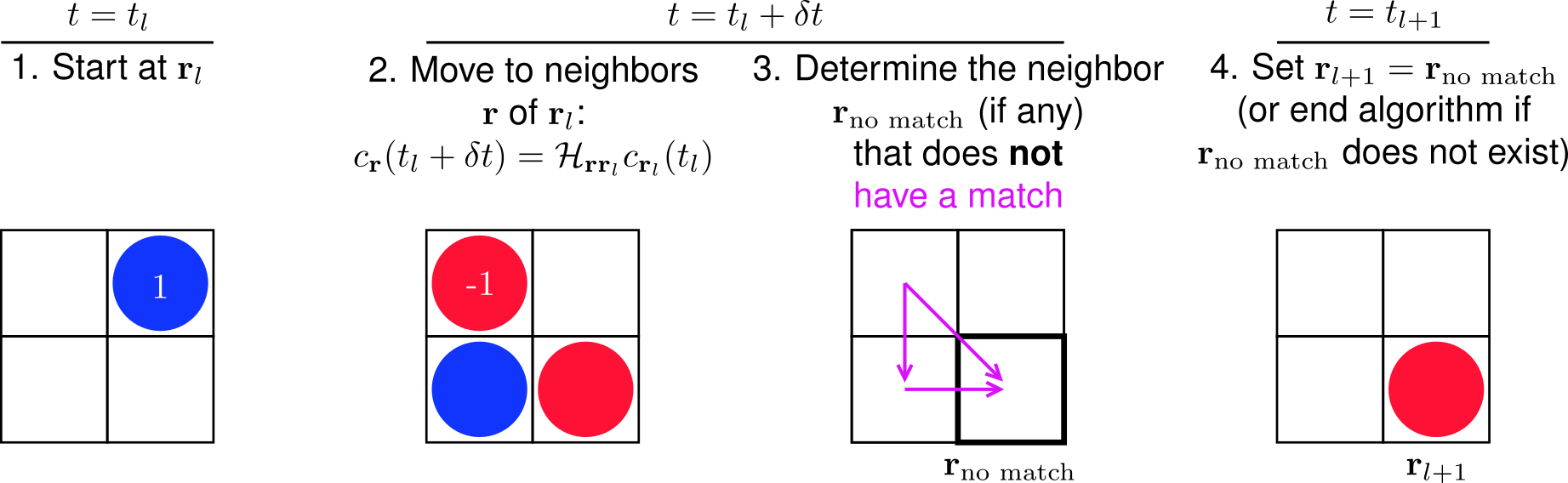}

\caption{\textbf{Algorithm 2.} Illustration of Algorithm 2.\label{fig:algorithm2}}

\end{figure}

Here are the steps of Algorithm 2 (Fig. \ref{fig:algorithm2}):
\begin{enumerate}
\item At the $l$th time step, $t=t_{l}$, the wavefunction is at site $\mathbf{r}_{l}$:
\begin{equation}
|\psi(t_{l})\rangle=c_{\mathbf{r}_{l}}'(t_{l})|\mathbf{r}_{l}\rangle=\pm|\mathbf{r}_{l}\rangle.\label{eq:psi_t_l2}
\end{equation}
\item Evolve the wavefunction forward by time $\delta t<t_{l+1}-t_{l}$:
\begin{equation}
|\psi(t_{l}+\delta t)\rangle=c_{\mathbf{r}_{l}}'(t_{l})\left(|\mathbf{r}_{l}\rangle+\sum_{\mathbf{r}\in\mathcal{N}(\mathbf{r}_{l})}\mathcal{H}_{\mathbf{r}\mathbf{r}_{l}}|\mathbf{r}\rangle\right).\label{eq:psi_t-p-dt2}
\end{equation}
\item Determine the neighbor $\mathbf{r}_{\text{no match}}$ (if any) of
$\mathbf{r}_{l}$ that does not \textit{match} with another neighbor
of $\mathbf{r}_{l}$. We say that $\mathbf{r}$ \textit{matches} with
$\mathbf{r}'$ if the probability amplitude of the former equals that
of the latter after multiplication by $\mathcal{H}_{\mathbf{r}'\mathbf{r}}$,
i.e., $\mathcal{H}_{\mathbf{r}'\mathbf{r}}c_{\mathbf{r}}'(t_{l}+\delta t)=c_{\mathbf{r}'}'(t_{l}+\delta t)$. 
\item If there is a neighbor $\mathbf{r}_{\text{no match}}$ of $\mathbf{r}_{l}$,
set
\begin{equation}
|\psi(t_{l+1})\rangle=c_{\mathbf{r}}'(t_{l}+\delta t)|\mathbf{r}_{\text{no match}}\rangle
\end{equation}
and $\mathbf{r}_{l+1}=\mathbf{r}_{\text{no match}}$; return to Step
2. If not, the algorithm terminates.
\end{enumerate}

\subsection{Derivation of Algorithm 2\label{subsec:algorithm2_derivation}}

In this section, we show that applying $f$ (Eq. \ref{eq:f}) to the
probability amplitudes $c_{\mathbf{r}}$ allows us to write each step
of Algorithm 1 as the same step of Algorithm 2. 

The derivation goes as follows:
\begin{enumerate}
\item By definition of $c_{\mathbf{r}_{l}}(t_{l})$ (Eq. \ref{eq:c_rl}),
\begin{align}
f\left(c_{\mathbf{r}_{l}}(t_{l})\right) & =\pm1.\label{eq:c_rl-map}
\end{align}
Comparing this equation to Eq. \ref{eq:psi_t_l2} for $|\psi(t_{l})\rangle$
of Algorithm 2, we see that applying $f$ to all $c_{\mathbf{r}}(t_{l})$
converts Step 1 of Algorithm 1 to the same step of Algorithm 2.
\item Notice that $H$ (Eq. \ref{eq:h}) has purely real NN couplings and
purely imaginary NNN couplings, i.e.,
\begin{equation}
H_{\mathbf{r}\mathbf{r}'}\in\begin{cases}
\mathbbm{R}, & \sigma(\mathbf{r})\text{ even, }\sigma(\mathbf{r}')\text{ odd},\\
\mathbbm{R}, & \sigma(\mathbf{r})\text{ odd, }\sigma(\mathbf{r}')\text{ even},\\
i\mathbbm{R}, & \sigma(\mathbf{r})\text{ even, }\sigma(\mathbf{r}')\text{ even},\\
i\mathbbm{R}, & \sigma(\mathbf{r})\text{ odd, }\sigma(\mathbf{r}')\text{ odd}.
\end{cases}\label{eq:h-parity}
\end{equation}
Using this equation, one can show that (see Eq. \ref{eq:psi_t-p-dt})
\begin{align}
c_{\mathbf{r}}(t_{l}+\Delta t) & =f\left(c_{\mathbf{r}_{l}}(t_{l})\right)\frac{\delta t}{\hbar}\times\begin{cases}
\text{Re}H_{\mathbf{r}\mathbf{r}_{l}}, & \sigma(\mathbf{r})\text{ even, }\sigma(\mathbf{r}_{l})\text{ odd},\\
-i\text{Re}H_{\mathbf{r}\mathbf{r}_{l}}, & \sigma(\mathbf{r})\text{ odd, }\sigma(\mathbf{r}_{l})\text{ even},\\
\text{Im}H_{\mathbf{r}\mathbf{r}_{l}}, & \sigma(\mathbf{r})\text{ even, }\sigma(\mathbf{r}_{l})\text{ even},\\
i\text{Im}H_{\mathbf{r}\mathbf{r}_{l}}, & \sigma(\mathbf{r})\text{ odd, }\sigma(\mathbf{r}_{l})\text{ odd}
\end{cases}\label{eq:c_r_t-p-dt}
\end{align}
for $\mathbf{r}\in\mathcal{N}(\mathbf{r}_{l})$. It follows that 
\begin{align}
f\left(c_{\mathbf{r}}(t_{l}+\delta t)\right) & =f\left(c_{\mathbf{r}_{l}}(t_{l})\right)\mathcal{H}_{\mathbf{r}\mathbf{r}_{l}},\quad\mathbf{r}\in\mathcal{N}(\mathbf{r}_{l}),\label{eq:c_r_t-p-dt_map}
\end{align}
where $\mathcal{H}$ is defined in Eq. \ref{eq:h_eff}. Moreover,
since $c_{\mathbf{r}_{l}}(t_{l}+\delta t)=c_{\mathbf{r}_{l}}(t_{l})$
(Eq. \ref{eq:psi_t-p-dt}), then 
\begin{align}
f\left(c_{\mathbf{r}_{l}}(t_{l}+\delta t)\right) & =f\left(c_{\mathbf{r}_{l}}(t_{l})\right).\label{eq:c_rl_t-p-dt_map}
\end{align}
Comparing Eqs. \ref{eq:c_r_t-p-dt_map}-\ref{eq:c_rl_t-p-dt_map}
to Eq. \ref{eq:psi_t-p-dt2} for $|\psi(t_{l})\rangle$ of Algorithm
2, we see that applying $f$ to all $c_{\mathbf{r}}(t_{l}+\delta t)$
converts Step 2 of Algorithm 1 to the same step of Algorithm 2.
\item Using Eq. \ref{eq:j-gen}, property \ref{eq:h-parity} of $H$, and
definition \ref{eq:h_eff} of $\mathcal{H}$, we can write the current
at time $t_{l}+\delta t$ from neighbor $\mathbf{r}$ of $\mathbf{r}_{l}$
to another neighbor $\mathbf{r}'$ of $\mathbf{r}_{l}$ as
\begin{align}
\langle J_{\mathbf{r}\rightarrow\mathbf{r}'}(t+\delta t)\rangle & =\frac{2(\delta t)^{2}}{\hbar^{3}}f\left(c_{\mathbf{r}'}(t+\delta t)\right)\mathcal{H}_{\mathbf{r}'\mathbf{r}}f\left(c_{\mathbf{r}}(t+\delta t)\right).
\end{align}
Since $\left|f\left(c_{\mathbf{r}'}(t+\delta t)\right)\right|=|\mathcal{H}_{\mathbf{r}'\mathbf{r}}|=\left|f\left(c_{\mathbf{r}}(t+\delta t)\right)\right|=1$,
the condition $\langle J_{\mathbf{r}\rightarrow\mathbf{r}'}(t+\delta t)\rangle>0$
is equivalent to $\frac{\hbar^{3}}{2(\delta t)^{2}}\langle J_{\mathbf{r}\rightarrow\mathbf{r}'}(t+\delta t)\rangle=1$,
or 
\begin{equation}
\mathcal{H}_{\mathbf{r}'\mathbf{r}}f\left(c_{\mathbf{r}}(t+\delta t)\right)=f\left(c_{\mathbf{r}'}(t+\delta t)\right).
\end{equation}
With this result and renaming $\mathbf{r}_{\text{receiver}}$ as $\mathbf{r}_{\text{no match}}$,
we can rewrite Step 3 of Algorithm 1 as the same step of Algorithm
2.
\item Using
\[
f\left(\text{sgn}\left[c_{\mathbf{r}}(t_{l}+\delta t)\right]\right)=f\left(c_{\mathbf{r}}(t_{l}+\delta t)\right)
\]
for all $\mathbf{r}$ and renaming $\mathbf{r}_{\text{receiver}}$
as $\mathbf{r}_{\text{no match}}$,we can rewrite Step 4 of Algorithm
1 as the same step of Algorithm 2.
\end{enumerate}
\clearpage{}

\noindent\textbf{Movie S1.} Dynamics generated by $H$ for the parameters
used in Fig. \ref{fig:dynamics}B. The probability of the system being
at each site is represented by a circle (area $\propto$ probability).\\
\\
\textbf{Movie S2.} Dynamics generated by $H$ for the parameters used
in Fig. \ref{fig:dynamics}C. The probability of the system being
at each site is represented by a circle (area $\propto$ probability).
\\
\\
\textbf{Movie S3.} Dynamics generated by $H$ for the parameters used
in Fig. \ref{fig:dynamics}D. The probability of the system being
at each site is represented by a circle (area $\propto$ probability).
\\
\\
\textbf{Movie S4.} Dynamics generated by $H$ for the parameters used
in Fig. \ref{fig:dynamics}E. The probability of the system being
at each site is represented by a circle (area $\propto$ probability).\\
\\
\textbf{Movie S5.} The dance where the initial dancers are on the
lattice edge. Snapshots are shown in Fig. \ref{fig:dance-dynamics}B.\\
\\
\textbf{Movie S6.} The dance where the initial dancers are on the
edge of a lattice with site defects. Snapshots are shown in Fig. \ref{fig:dance-dynamics}C.\\
\\
\textbf{Movie S7.} The dance where the initial dancer is on the inner
edge of a lattice with a hole in the middle (i.e., Corbino geometry).
Snapshots are shown in Fig. \ref{fig:dance-dynamics}D.\\
\\
\textbf{Movie S8.} The dance where the initial dancer is in the lattice
bulk. Snapshots are shown in Fig. \ref{fig:dance-dynamics}E.